\documentstyle[procl,epsfig]{article}

\input{epsf}

\newcommand{\fc}{$ F_{2}^{c\overline{c}}\,$}

\newcommand{\GV}{GeV}
\newcommand{\GVsq}{GeV$^2$}

\def\gsim{\mathrel{\rlap{\lower4pt\hbox{\hskip1pt$\sim$}}
    \raise1pt\hbox{$>$}}}         

\def\frac#1#2{{{#1}\over {#2}}}

\def\smallfrac#1#2{\hbox{${{#1}\over {#2}}$}}

\def\MS{\hbox{$\overline{\rm MS}$}}

\catcode`@=11 
\def\slash#1{\mathord{\mathpalette\c@ncel#1}}
 \def\c@ncel#1#2{\ooalign{$\hfil#1\mkern1mu/\hfil$\crcr$#1#2$}}
\def\lsim{\mathrel{\mathpalette\@versim<}}
\def\gsim{\mathrel{\mathpalette\@versim>}}
 \def\@versim#1#2{\lower0.2ex\vbox{\baselineskip\z@skip\lineskip\z@skip
       \lineskiplimit\z@\ialign{$\m@th#1\hfil##$\crcr#2\crcr\sim\crcr}}}
\catcode`@=12 

\def\PR{{\it Phys.~Rev.~}}

\def\NP{{\it Nucl.~Phys.~}}
\def\NPBPS{{\it Nucl.~Phys.~B (Proc.~Suppl.)~}}
\def\PL{{\it Phys.~Lett.~}}

\def\ZP{{\it Zeit.~Phys.~}}

\def\vol#1{{\bf #1}}\def\vyp#1#2#3{\vol{#1}, #3 (#2)}

\def\be{\begin{equation}}
\def\ee{\end{equation}}
\def\bea{\begin{eqnarray}}
\def\eea{\end{eqnarray}}

\begin{document}

\pagestyle{empty}

\begin{flushright}
{\tt hep-ph/9609309}\\
{DESY 96-185}\\
{Edinburgh 96/24}\\
\end{flushright}
\vskip 12pt
\begin{center}
{\bf STRUCTURE FUNCTIONS} 
\vskip 24pt
{Richard D. Ball\footnote[1]{Royal Society University Research Fellow}}\\
\vskip 12pt
{\em Department of Physics and Astronomy}\\
{\em University of Edinburgh, EH9 3JZ, Scotland}\\
\vskip 18pt
{and}
\vskip 18pt
{Albert De Roeck}\\
\vskip 12pt
{\em DESY, Notkestr.85, }\\ 
{\em D-22607 Hamburg, Germany}\\ 
\vskip 36pt
\end{center}
\abstracts{
We summarize recent developments in the understanding of nucleon structure. 
New data on $F_2$, $R$ and \fc, over a wide range of $Q^2$
(from $10^4$\GVsq\ down to $0.1$\GVsq) and $x$ (down to $10^{-6}$),
are described. Conventional leading twist NLO perturbative
QCD gives an excellent description of all the new data with $Q^2$ above 
a \GV, leaving very little room for either higher twists or higher logarithms.
We summarize the current status of NLO fits, and the determination of 
the gluon distribution and the strong coupling constant from structure 
function data. Finally we consider some of the theoretical issues raised
by the new data.
}
\smallskip
\begin{center}
{Summary Talk of WG1 at {\em DIS96}, Rome, April 1996}\\
\smallskip
{\em to be published in the proceedings}\\
\end{center}
\bigskip
\begin{flushleft}
{August 1996}
\end{flushleft}
\eject

\pagenumbering{arabic}
\pagestyle{plain}

\title{STRUCTURE FUNCTIONS}

\author{ R.D. BALL}
\address{Department of Physics and Astronomy, University of Edinburgh,\\
Edinburgh EH9 3JZ, Scotland}

\author{ A. DE ROECK}
\address{ DESY, Notkestr.85, D-22607 Hamburg, Germany }

\maketitle\abstracts{
We summarize recent developments in the understanding of nucleon structure. 
New data on $F_2$, $R$ and \fc, over a wide range of $Q^2$
(from $10^4$\GVsq\ down to $0.1$\GVsq) and $x$ (down to $10^{-6}$),
are described. Conventional leading twist NLO perturbative
QCD gives an excellent description of all the new data with $Q^2$ above 
a \GV, leaving very little room for either higher twists or higher logarithms.
We summarize the current status of NLO fits, and the determination of 
the gluon distribution and the strong coupling constant from structure 
function data. Finally we consider some of the theoretical questions issues
by the new data.
}

Until a few years ago our knowledge of structure functions and
derived quantities such as parton distributions came almost entirely
from fixed-target experiments, using muon and electron 
beams. Now it is being complemented
and extended by results from the HERA $ep$ collider, especially
in the regions of low $x$ and high $Q^2$. 
At this workshop both the fixed target and the 
HERA experiments have 
presented new data on $F_2$, extending the kinematical range and 
reaching a precision of more than a factor two better than the 
results from the 1993 run shown at last year's meeting in Paris.\cite{Paris}
The ZEUS data now reach $Q^2$ values down to 0.1 \GVsq, and $x$ values 
as low as $10^{-6}$. New measurements of the ratio of the 
longitudinal to transverse photon absorption cross section $R$ were  
shown by NMC and CCFR. A first glimpse of $F_L$ at low $x$ was
presented by H1. There were also new data on the charm quark 
contribution to $F_2$.

Already at last year's meeting it was clear that HERA data at low $x$ 
but with $Q^2$ above a few \GVsq\ are well described by conventional 
leading twist NLO perturbative QCD evolution from a flat or 
valence-like boundary condition. The increased precision of the new 
data have considerably reinforced this conclusion, putting  
tight constraints on the size of novel effects such as BFKL 
logarithms (i.e. higher orders) and parton recombination (i.e. 
higher twist). The apparent absence of such contributions poses various
interesting theoretical problems, some of which were discussed at this
meeting.

On the other hand the serendipitous success of NLO perturbative QCD 
at small $x$ encourages us to use it to extract parton distributions 
in the proton, and in particular the gluon content of the proton at 
low $x$. At large $x$ the gluon can be constrained  with the help 
of the newly available inclusive jet data from the Tevatron and an 
improved understanding of the older prompt photon data.
Both of these recent developments lead to new information on the 
strong coupling.

Here we only discuss in detail data on unpolarized nucleon structure 
functions: data on diffractive and polarized nucleon 
structure functions was presented in the sessions 
of other working groups, and discussed in joint sessions.

\begin{figure}[htbp] \centering        
\epsfig{file=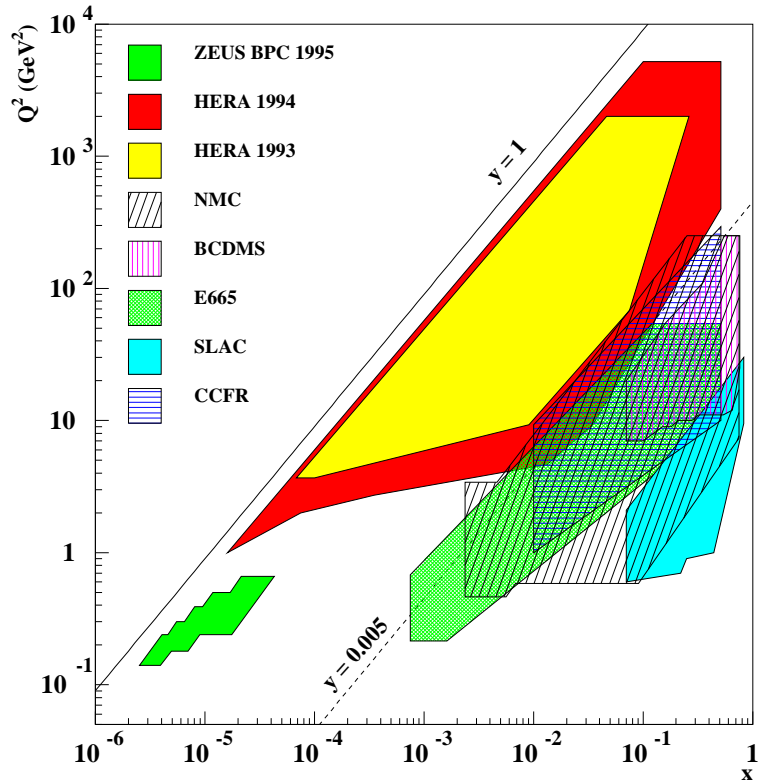,width=100mm}
\caption[]{ The kinematic region covered by proton structure 
function data, including the new measurements shown at this 
workshop.}
\label{fig1}
\end{figure}                      

\section{The Structure Function Data}

The differential cross section for neutral current deep inelastic 
scattering is related to the three structure functions
$F_2, R=F_2/2xF_1$ and $F_3$ according to
\begin{equation}              
  \frac{d^2\sigma^{ep\rightarrow eX} }
{dx dQ^2} =\frac{4\pi\alpha^2}{Q^4x}
    \left[\big(1-y+\frac{y^2}{2(1+R)}\big)F_2(x,Q^2)
                               \mp y(1-y)xF_3(x,Q^2)\right]. 
\label{dsigma}                         
\end{equation}
Here $Q^2$ is the four momentum transfer, $x$ the Bjorken-$x$ value and 
$y$ the energy transfer from the electron to the proton in the proton
rest frame. The parity violating structure function $F_3$ only becomes 
significant in the region where $Z^0$ exchange or $Z^0-\gamma$ interference 
dominates, i.e. at large $Q^2$. The effect of $R$ is important at large $y$, 
typically $y\gsim 0.4$.

The new kinematical region covered by the HERA and fixed target data
is shown in fig.\ref{fig1}. The HERA data~\cite{bassler,kooijman}
were collected in 1994 and 
represent a tenfold increase in statistics with respect to the 1993 
data, allowing a significant extension of 
the covered region towards higher $Q^2$. A further extension 
towards higher $x$ was possible due to better 
methods of calculation of the kinematic
variables from the hadronic final states.\cite{F294H1,F294ZEUS}
The improvement is due to using a mixture of the electron and
hadron information, and momentum balance.
For not too high $Q^2$ the ZEUS data reach $y$ values as low as 
$y\sim 0.005$, hence
the HERA data have now a small region of overlap 
with data from fixed target experiments.
Such an overlap will be important for precise gluon density 
extractions.
A substantial extension to smaller $Q^2$ was achieved by shifting the
interaction vertex for part of the time, by analysing events with a 
hard photon radiated in the initial state, and by improving the detector
components around the beampipe. This allowed both H1 and ZEUS to reach
$Q^2$ values down to about $1$ \GVsq.

\begin{figure}[thbp] \centering        
\epsfig{file=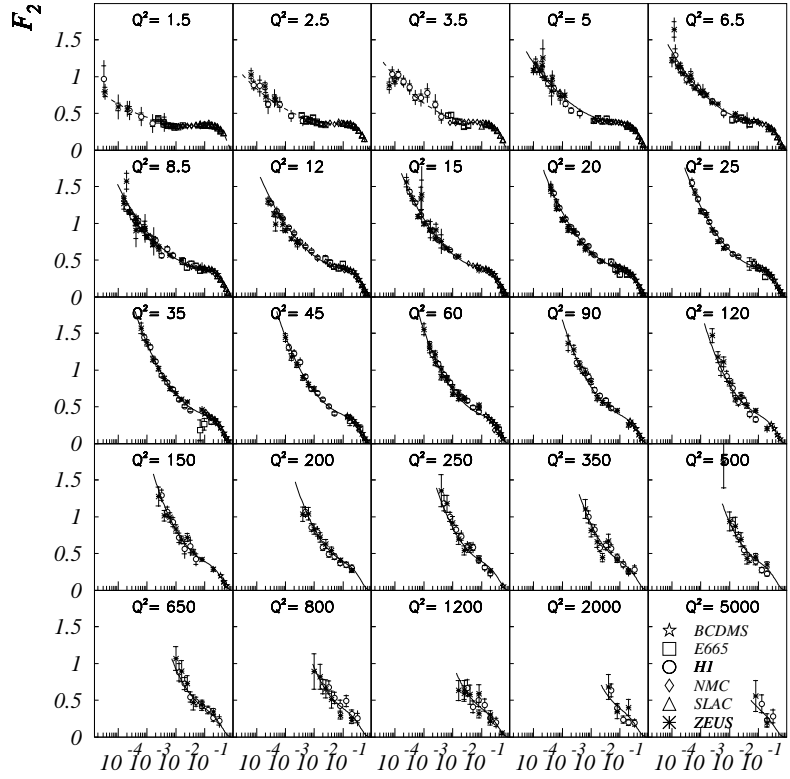,width=12cm,
bbllx=0pt,bblly=100pt,bburx=600pt,bbury=650pt}
\vskip-1cm
\caption[]{ The structure function $F_2$ as function of $x$ for 
different $Q^2$ values, including
measurements
from H1, ZEUS, NMC, E665, SLAC and BCDMS. The solid line is a NLO QCD fit
made using data with $Q^2> 5$ GeV$^2$,  
from 
H1, BCDMS and NMC.}
\label{fig2a}
\end{figure}

\begin{figure}[htbp] \centering        
\epsfig{file=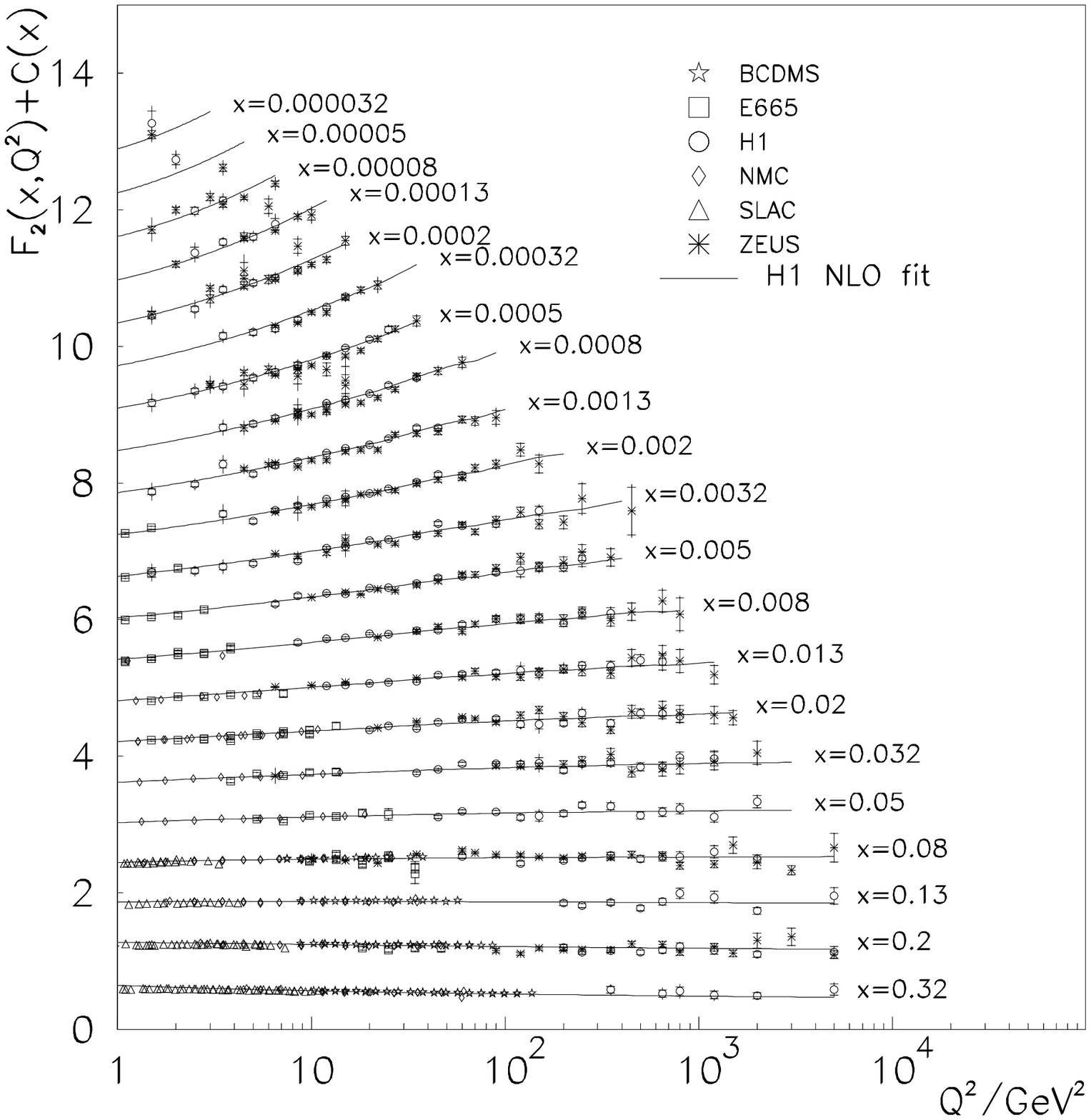,width=12cm,
bbllx=0pt,bblly=100pt,bburx=600pt,bbury=650pt}
\vskip-1cm
\caption[]{ The structure function $F_2$ as function of $Q^2$ for 
different $x$ values, including
measurements
from H1, ZEUS, NMC, E665, SLAC and BCDMS. The solid line is a NLO QCD fit
made using data with $Q^2> 5$ GeV$^2$,  
from 
H1, BCDMS and NMC.}
\label{fig2}
\end{figure}

\begin{figure}[htbp]
\begin{center}
\epsfig{file=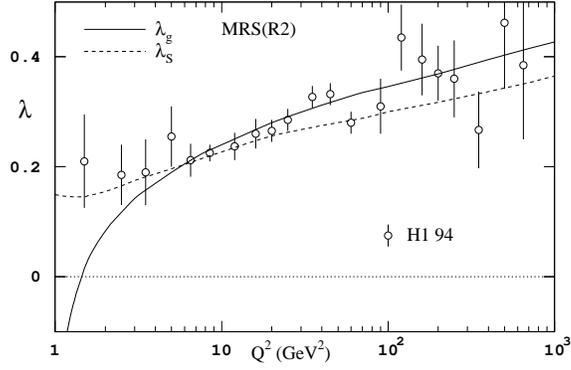,width=10cm,bbllx=0pt,bblly=350pt,bburx=600pt,bbury=700pt}
\end{center}
\vskip-1cm
\caption[]{\label{fig3} 
The exponent $\lambda$ as measured by the H1 collaboration\cite{F294H1} 
using fits of the form $F_2 \sim x^{-\lambda}$ at fixed $Q^2$
values and for $x< 0.1$. Also shown are $\lambda_S$ and $\lambda_g$ 
as a function of $Q^2$ calculated for the MRS R2 
parametrizations.\cite{roberts} 
A similar plot extracted using a rather different technique  
may be found in ref.30.
}
\end{figure}

\begin{figure}[htbp]
\begin{center}
\epsfig{file=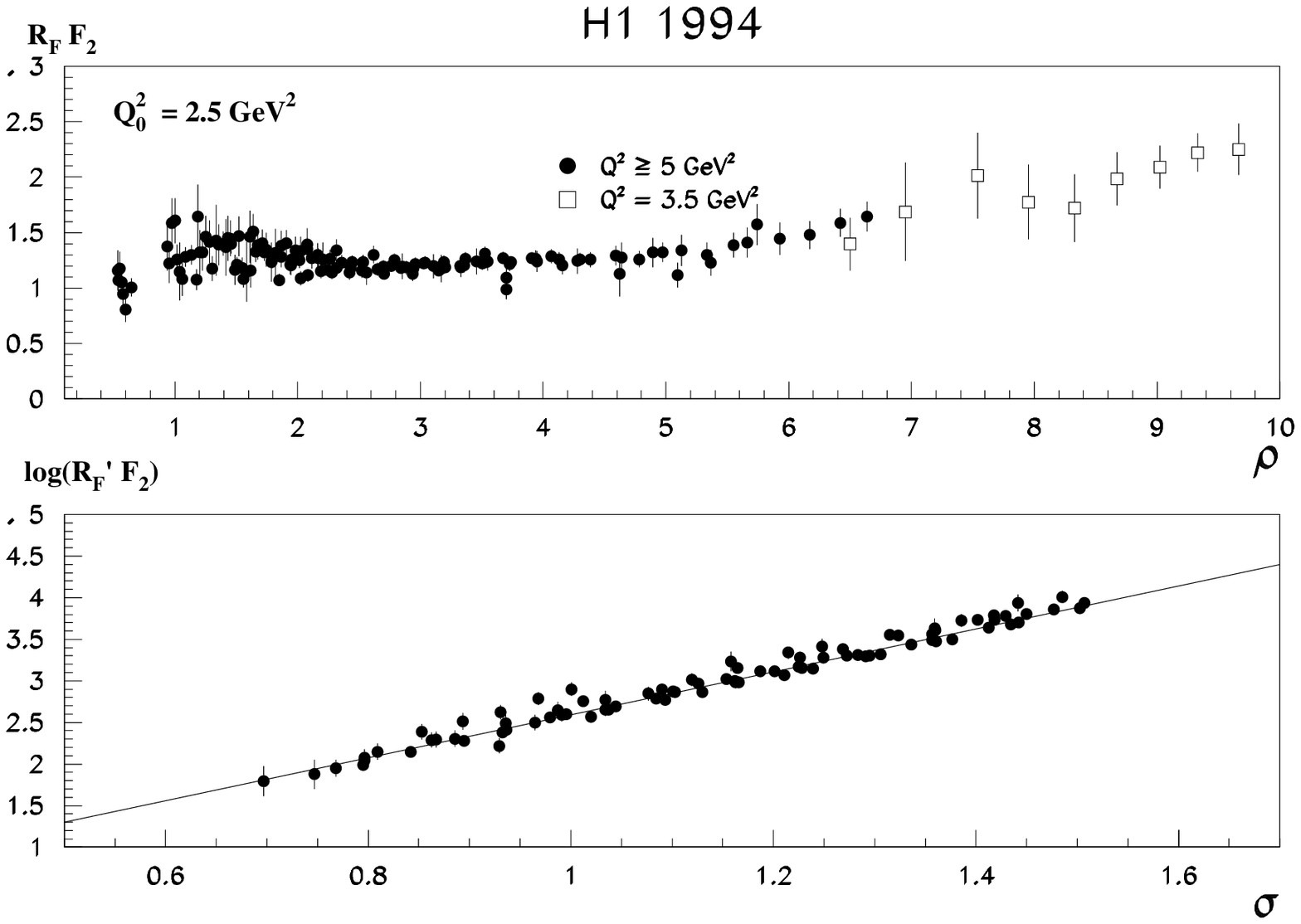,width=8cm,bbllx=35pt,bblly=200pt,bburx=540pt,bbury=610pt}
\end{center}
\vskip-0.5cm
\caption[]{\label{fig4}
The rescaled structure functions $R_FF_2$ versus $\rho$
and $\log(R_F'F_2)$ versus $\sigma$ (see text).\cite{F294H1}
Only data with $Q^2\ge$ 5  GeV$^2$ and $\rho>$ 2 are shown
in the second figure. The rescaling functions 
$R_F$ and $R_F'$, and $\alpha_s(Q)$ in the definition (2), 
are all evaluated using the two loop expressions in ref.12.}
\end{figure}

\subsection{Measurements of $F_2$}

The $F_2^p$ data from the HERA 1994 run,\cite{F294H1,F294ZEUS} 
together with the published 
data from fixed target experiments,\cite{nmc,bcdms,e665,slac}
are shown in figs.\ref{fig2a} and \ref{fig2}. Good agreement 
between the two HERA
experiments is seen, as well as a smooth continuation from the
fixed target data to the collider data.
At fixed $Q^2$ $F_2$ rises with decreasing $x$ down to the smallest 
values of $Q^2$, although the steepness of the rise is clearly seen
to decrease with decreasing $Q^2$. Similarly at fixed $x\lsim 0.1$ 
$F_2$ rises with $Q^2$, the rise becoming steeper as $x$ decreases. 

The evolution of the rise with $x$  may be demonstrated by 
parameterizing $F_2 \sim x^{-\lambda}$ at fixed $Q^2$ values.
The result for $\lambda$ as function of $Q^2$ is shown in fig.\ref{fig3}.
The parameter $\lambda$ clearly becomes smaller with decreasing $Q^2$ 
(although the region in $x$ over which the function $ x^{-\lambda}$
is fitted is reduced for each $Q^2$ bin, which slightly enhances the 
effect). 
 
These two rises, with decreasing $x$ (fig.\ref{fig2a}) and increasing $Q^2$
(fig.\ref{fig2}) may be neatly combined together: at large enough $Q^2$ 
the data in the low $x$ region display the universal non-Regge rise 
from a flat boundary condition predicted by perturbative QCD.\cite{DGPTWZ,das}
This is demonstrated in fig.~\ref{fig4},\cite{F294H1} where the rescaled $F_2$ 
is shown as function of the two scaling variables 
\begin{equation}
\sigma \equiv 
\sqrt{\log(\smallfrac{x_0}{x})\log(\smallfrac{\alpha_s(Q_0)}{\alpha_s(Q)})},  
\qquad
\rho \equiv \sqrt{\log(\smallfrac{x_0}{x})\big/
\log(\smallfrac{\alpha_s(Q_0)}{\alpha_s(Q)})};
\end{equation}
the rescaling factor $R_F'$ is a simply calculable subasymptotic 
function, while
$R_F=e^{-2\gamma\sigma}R_F'$. The measured slope of the rise
with $\sigma$ is $2.57\pm 0.08$, to be compared with the QCD prediction of 
$2\gamma = 2.5$ for five active flavors. There exist other satisfactory 
parameterizations the data,\cite{haidt,naumann} but these do not predict the 
slope of the rise with $x$ and $Q^2$.
 
An excellent fit to the data may be achieved using the full NLO 
perturbative QCD evolution equations, with a fitted boundary condition,
as demonstrated in figs.\ref{fig2a} and \ref{fig2}. More details 
of such fits will be 
discussed in the next section. Models based on Regge phenomenology, 
but not including QCD evolution, generally undershoot the data even 
for the smallest $Q^2$ values shown.

The increase in statistics means that it is possible to make  
significant differential measurements in the high $Q^2$ region for 
the first time.\cite{williams,pieuchot} The data are beginning to become 
sensitive to the valence and sea quark distributions, but it is 
as yet too early for significant measurements of $F_3$.
Combining the data on charged current interactions of both experiments
a measurement on the $W$ propagator mass yields $M_W=82^{+6+3}_{-5-3}$ GeV,
consistent with direct measurements at the proton-antiproton colliders.
The large $Q^2$ end of the $d\sigma/dQ^2$ spectrum for  neutral current data
is used to set limits on the effective scale $\Lambda$ for contact 
interactions. Depending on the sign of the interference term with the 
Standard Model interactions, 95\% C.L. limits lead to lower limits of 
$\Lambda$ in the range 1.0-2.5 TeV.

\begin{figure}[htbp] 
\centering        
\epsfig{file=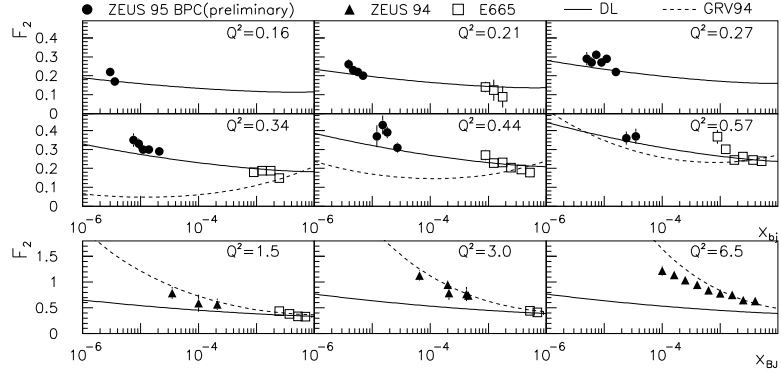,bbllx=15pt,bblly=0pt,bburx=630pt,bbury=350pt,width=10cm}
\vskip-0.5cm
\caption[]{ Preliminary measurement of the structure function $F_2$ 
as function of $x$ for different $Q^2$ values from ZEUS, 
with the new beampipe calorimeter (full circles), and from the  nominal 
ZEUS analysis (full triangles) and E665 (open squares). 
The data analysis assumes that $F_L = 0$. A 5\% normalization 
error is not shown. Predictions for 
the \hbox{Donnachie-Landshoff\protect\cite{DL} (full)} and 
\hbox{GRV\protect\cite{GRV} (broken)} predictions are given as well.}
\label{fig5}
\end{figure}                 

\begin{figure}[htbp]
\begin{center}
\epsfig{file=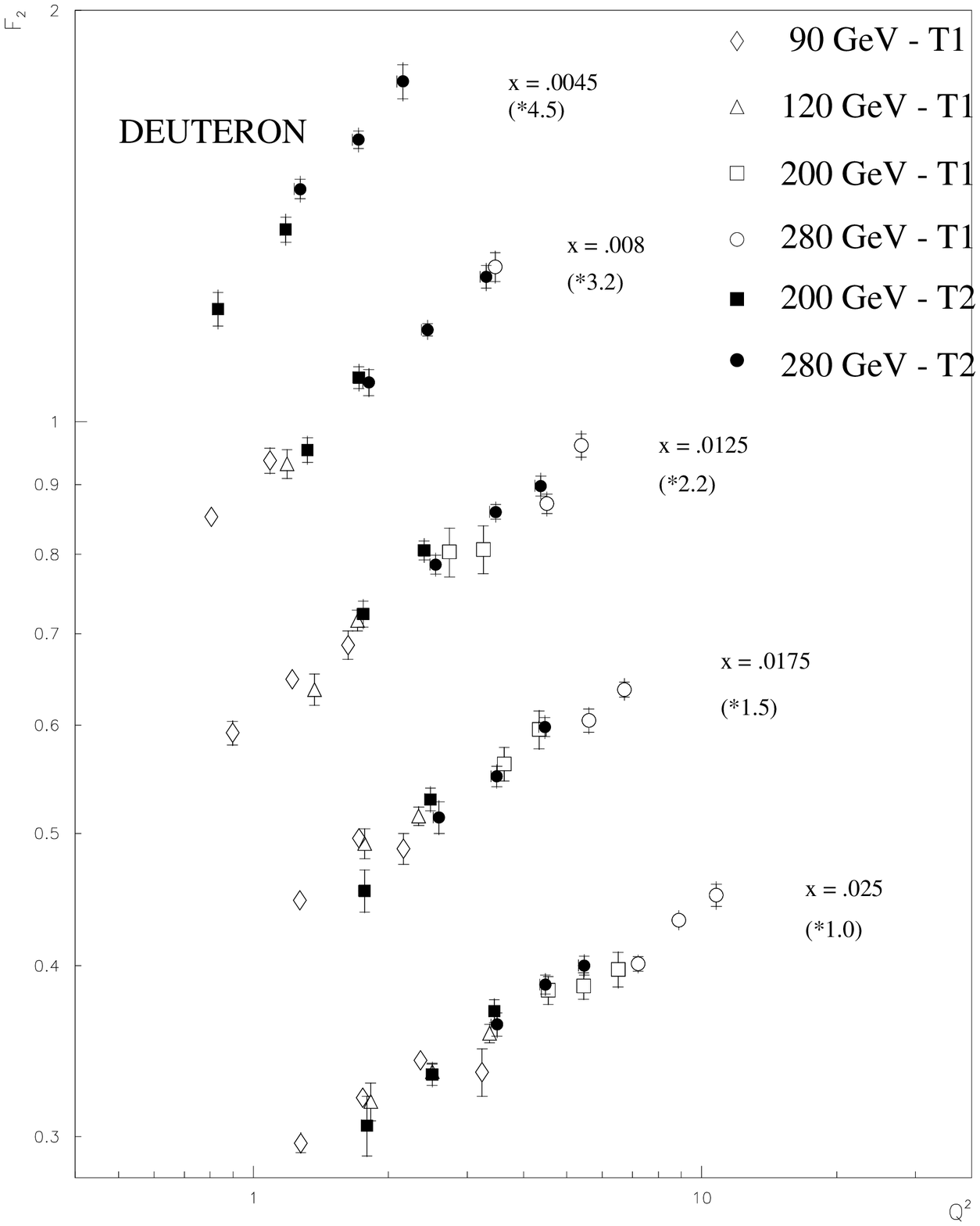,width=4.8cm,
bbllx=15pt,bblly=145pt,bburx=600pt,bbury=650pt} 
\epsfig{file=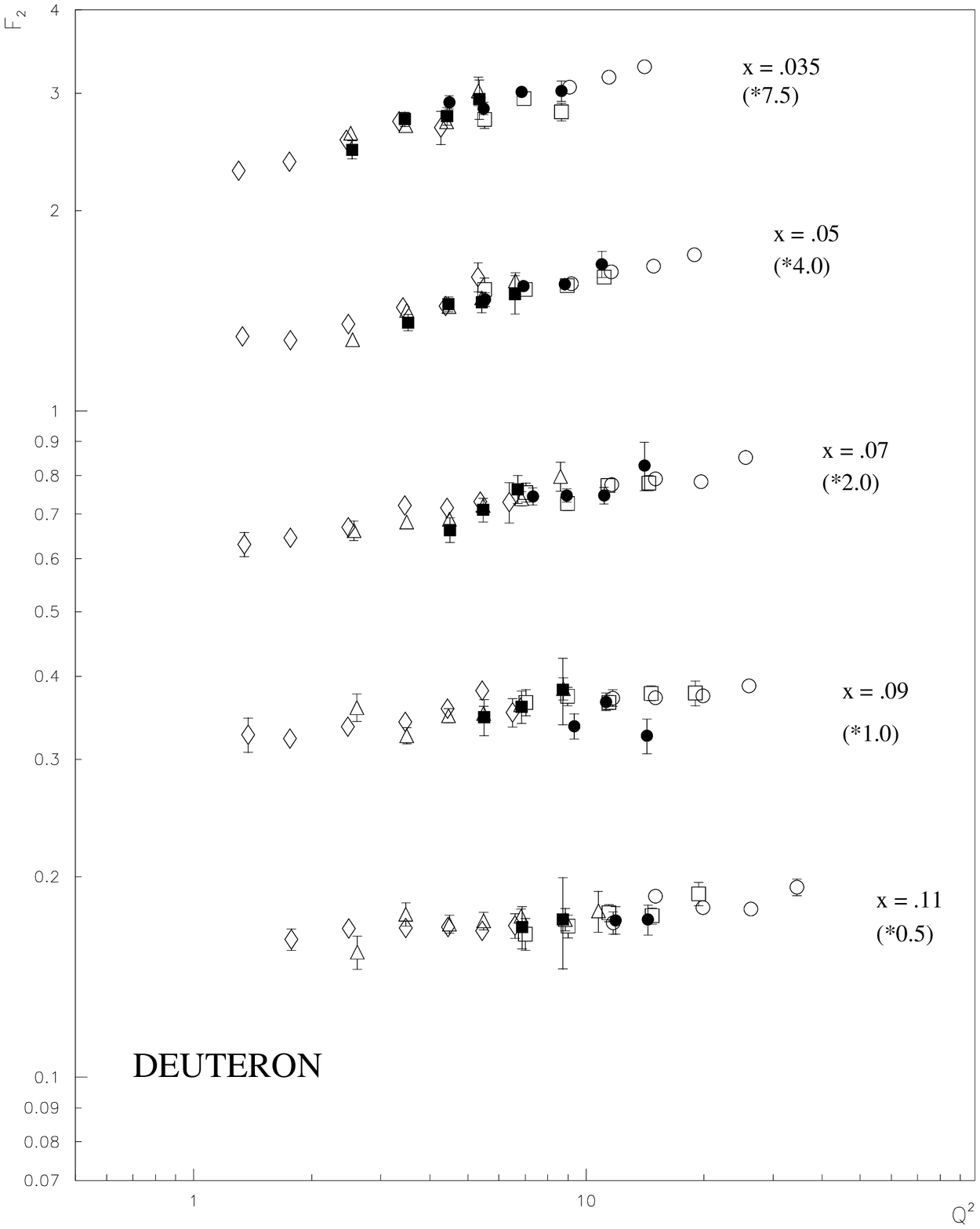,width=4.8cm,
bbllx=15pt,bblly=145pt,bburx=600pt,bbury=650pt}
\end{center}
\vskip 0.5cm
\caption[]{\label{fig6}
Comparison of the results at different energies and triggers for the 
deuteron structure function $F^d_2$ from the NMC.\cite{kabuss} The 
full symbols are the new (preliminary) data points. The inner 
error bars are statistical, the full ones represent the total error.}
\end{figure}

First preliminary results were presented by ZEUS of data taken with a
small calorimeter placed 3m downstream of the interaction point to
accept electrons deflected under a very small angle.\cite{zhu}
The kinematical region covered is $0.16<Q^2<0.57$ \GVsq\ and 
$3.10^{-6}<x<3.10^{-5}$. The data are shown in 
fig.~\ref{fig5} together with data
from the muon experiment 
E665.\cite{e665} The latter  cover the region
$0.0008<x<0.06$ and $0.2<Q^2<75$ \GVsq. While the  
Regge inspired model\cite{DL} predicts too low a cross section at low $x$ for 
$Q^2$ values above 1 \GVsq, the agreement 
in this region is much better. It thus appears that there is a transition 
from the perturbative region to the Regge region at
$Q^2\sim 1$ \GVsq. Future HERA data 
will cover this region and will thus bridge the two presently 
disjoint data samples.

New preliminary results on $F_2$ from muon-proton and muon-deuteron
scattering have been shown by the NMC collaboration.\cite{kabuss}
The data taken with the small angle trigger in 1989 
have been analysed, giving access to a lower $x$ and $Q^2$ region than in 
previous NMC measurements. For incident muon beam energies of 200 and 280
\GV, scattered muons were detected with angles down 
to 6 mrad. The proton and deuteron structure functions were measured
in the $x$ range from 0.0045 to 0.6 and the $Q^2$ range of
0.5 to 75 \GVsq. The $F_2$ for deuterons is shown in fig.~\ref{fig6}
including the old and new datasets.
The new NMC results compare very well with the E665 results.
For the ratio of
$F_2$ for neutrons to protons, the range covered extends to $0.001<x<0.8$
and $0.1<Q^2<145$ \GVsq. 
The $Q^2$ dependence of this ratio shows slopes which are larger than  
expected from QCD at large $x$, while there is essentially no 
$Q^2$ dependence measured a small $x$. The deuteron data show indications 
of a small amount of shadowing at small $x$.

\subsection{Measurements of $R$}

New preliminary data on $R = \sigma_L/\sigma_T$ have been presented from the 
neutrino experiment CCFR,\cite{bodek} from NMC\cite{milst} and from 
H1.\cite{klein}
The CCFR data cover values of $x$ in the range $0.01<x<0.6$
for the range 
$4<Q^2<300$ \GVsq. The new 
data are in agreement with the SLAC fit through previous data.
NMC has released new data on $R$ measured on proton and deuterium targets, 
for $x$ values in the range $0.002 <x<0.12$ and $1.5< Q^2<20$ \GVsq.
The systematic errors are preliminary 
around 0.1, flat in $x$, and are dominated by the 0.15\% normalization 
uncertainty between the two datasets. The difference $R^p$-$R^d$ is consistent 
with zero.

Despite the availability of data 
on $R$ for $x$ below 0.1 from fixed target experiments, 
a substantial uncertainty still exists in the extrapolation of $R$ 
down to the low-$x$ region at HERA.
Furthermore at small $x$ the quantity $R$ is a sensitive probe of the gluon
density in the proton.
Therefore measurements of $R$ at HERA are mandatory.

\begin{figure}[htbp] 
\centering        
\epsfig{file=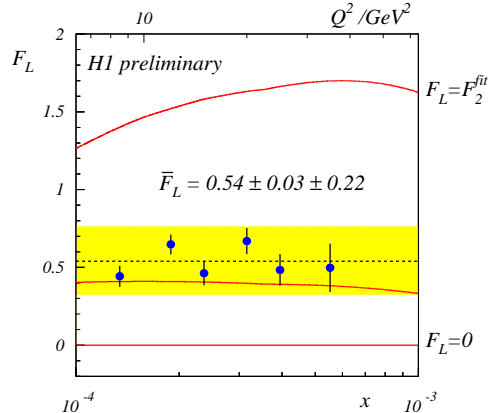,bbllx=0pt,bblly=150pt,bburx=600pt,bbury=650pt,width=8cm}
\vskip-1cm
\caption[]{\label{fig7} 
Measured longitudinal structure function $F_L$ by H1\cite{klein} 
for $y=0.7$ as functions of $x$ and $Q^2$ for 
$Q^2=$ 8.5, 12, 15, 20, 25, and 35 \GVsq. The error bars are 
statistical errors, while the band 
is the systematic error common to all points. The full line is a perturbative 
calculation of $F_L$ using the quark and gluon 
distributions as determined by the NLO QCD fit to the data.}
\end{figure}                      

While $R$ can be measured at fixed target experiments by varying the
incoming beam energy (which results in a variation of $y$ for 
fixed $x$ and $Q^2$
in expression (\ref{dsigma}), no such opportunity exists so far at HERA.
However a first glimpse of $R$ at small $x$ can be obtained by 
taking the measurement
of $F_2$ in a region where the effect of $R$ is negligible (for 
example $y< 0.35$), making a NLO QCD fit to these data, and then using 
the fit to extrapolate into the region of large $y$ (for example $y = 0.7$). 
The comparison of the extrapolated value of $F_2$ with the 
experimentally measured cross section then gives a handle on $R$. 
The result is shown in fig.\ref{fig7}.
For this determination the cross section measurement has been extended to 
$y=0.7$. The value of $F_L = F_2-2xF_1$ is around 0.5 with a systematic 
error of about 0.2. The result is self-consistent, in the sense that it 
is compatible with the QCD prediction obtained using the gluon 
distribution found in the fit.

Future prospects for the measurement of $R$ at HERA by reducing the
proton beam energy were also discussed.\cite{bauerdick} A precision 
of about 0.3 can be reached on $R$ using two beam energies 
(820 and 450 GeV) with the 1996 luminosity. The measurement is again 
limited by systematics.

\subsection{Measurements of \fc}

\begin{figure}[thbp]
\begin{center}
\epsfig{file=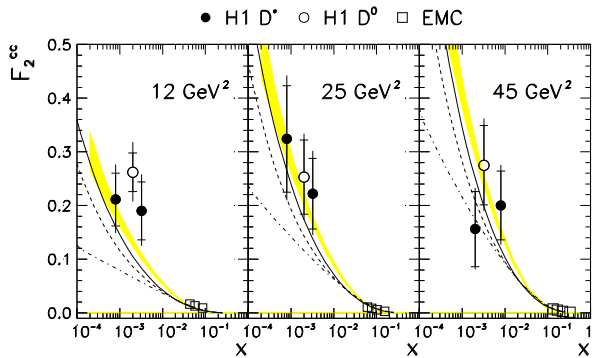,width=8cm,
bbllx=30pt,bblly=30pt,bburx=530pt,bbury=470pt}
\end{center}
\vskip-1.5cm
\caption[]{\label{fig8}
$F_2^{c\overline{c}}$ as derived from the $D^{*+}$ (full dots) and the 
$D^0$ (open circles) cross sections measured by H1. Statistical 
(thick) and total (thin) errors are shown. The EMC data 
is also included (open squares).
The data are compared with NLO predictions using GRV (full line),
MRSH (dashed line) and MRSD0' (dash-dotted line) for $m_c = 1.5 $\GV.
The shaded band represents the prediction from the H1 NLO QCD fit to the 
$F_2$ measurements.}
\end{figure}

Also reported at this meeting were results on the charm contribution
to $F_2^p$, \fc. H1~\cite{daum} and ZEUS~\cite{cassigari} have shown 
that deep inelastic scattering 
events with charm can be tagged through the detection of $D^0$ 
and $D^{*\pm}$ mesons. The decays $D^0\rightarrow K^-\pi^+$ and 
$D^{*+}\rightarrow D^0\pi^+\rightarrow K^-\pi^+\pi^+$ 
(and charge conjugate channels) have been analysed. 
H1 also demonstrated that the production mechanism for 
charm in DIS is compatible with photon-gluon fusion, the 
non-perturbative
charm sea contribution being no more than a few per cent. Hence \fc\ 
may be useful as a direct probe of the gluon density in the proton. 
The results of the \fc\ measurements, assuming $R=0$, both for $D^0$ and $D^*$ 
meson production, are displayed in fig.~\ref{fig8}. 
The errors refer to the statistics and to the experimental 
systematics. These measurements extend previous knowledge of  
\fc\ (from measurements by the EMC experiment~\cite{emc}) towards 
smaller values of $x$ by two orders of magnitude.  
When combined with this earlier measurement a steep rise 
of \fc\ is observed with decreasing $x$. This is consistent with
predictions from NLO fits.

\vskip-0.5cm
 
\section{Parton Distributions and the Strong Coupling}

The success of conventional leading twist NLO perturbative QCD in 
describing all available structure function data with $Q^2\gsim 1$ \GVsq, 
even that at very small $x$, means that we can use the data to extract 
input parton distributions (and in particular the gluon distribution) 
and the strong coupling constant. At this 
meeting new global fits were presented by both the CTEQ group\cite{tung}
(the CTEQ4 distributions) and the MRS group\cite{roberts} (the R 
distributions). While the CTEQ starting scale $Q_0^2$ generally remains 
at $2.6$ \GVsq, they have in the light of recent data also offered a 
distribution with $Q_0^2=0.7$ \GVsq; MRS have now reduced the starting scale 
of all their latest distributions from $4$ \GVsq\ to $1$ \GVsq. Both groups 
now assume that at small $x$ the 
input singlet quark and gluon distributions behave as $x^{-\lambda_S}$,
$x^{-\lambda_g}$ respectively, the two parameters $\lambda_S$ and 
$\lambda_g$ being now fitted independently, rather than assumed to be equal.
Note that these `effective powers' must be interpreted with some care: their
precise values are dependent on the form of the parameterization, on
the choice of scheme (the data in fig.\ref{fig3} are by construction in 
DIS scheme, while the $\lambda_S$ curve is in \MS) and the treatment 
of thresholds. Nonetheless, certain qualitative conclusions may be drawn:
both powers increase with $Q^2$, $\lambda_S$ in 
line with that extracted from the H1 data (see fig.\ref{fig3}), 
and $\lambda_g$ leading $\lambda_S$. At smaller scales however, 
somewhere between $1$ and $4$ \GVsq, $\lambda_g$ goes negative and 
the gluon becomes increasingly valence-like. 

The soft (i.e. flat or valence-like) initial gluon found in these fits is of 
course perfectly consistent with the double asymptotic scaling seen in 
fig.\ref{fig4},
since this characterizes the non-Regge rise\cite{DGPTWZ} generated 
dynamically in perturbative QCD by gluon bremsstrahlung from a flat input.
It also accounts for the remarkable success of the GRV\cite{GRVpred} 
prediction of this rise, in which valence-like 
inputs\cite{chyla} were evolved 
from a very low scale $Q_0^2=0.34$ \GVsq: the shape of any rise 
generated dynamically from 
a valence-like input at a very low scale is universal (while its 
overall normalization, not calculable in perturbation theory, was fixed 
by GRV by comparison with prompt photon data, and imposition of the 
momentum sum rule).\cite{vogtp} 

The non-Regge rise\cite{DGPTWZ} of the parton 
distributions is actually slower than any power of $x$, while faster 
than any power of $\ln 1/x$, growing instead as a power of 
$\exp\sqrt{\ln 1/x}$. In fact the data are not consistent with 
perturbative evolution from input distributions at large $Q_0^2$ 
which rise as a power of $x$, since the growth with $Q^2$ of the 
slope of the rise is no longer reproduced\cite{das}   
(although in any particular $Q^2$ bin it is not 
yet easy\cite{haidt} to distinguish between a power of $x$ and a power 
of $\ln 1/x$). It is thus useful to consider fits 
with alternative parameterizations of the small-$x$ input.\cite{zomer}

\begin{figure}[htbp]
\begin{center}
\epsfig{file=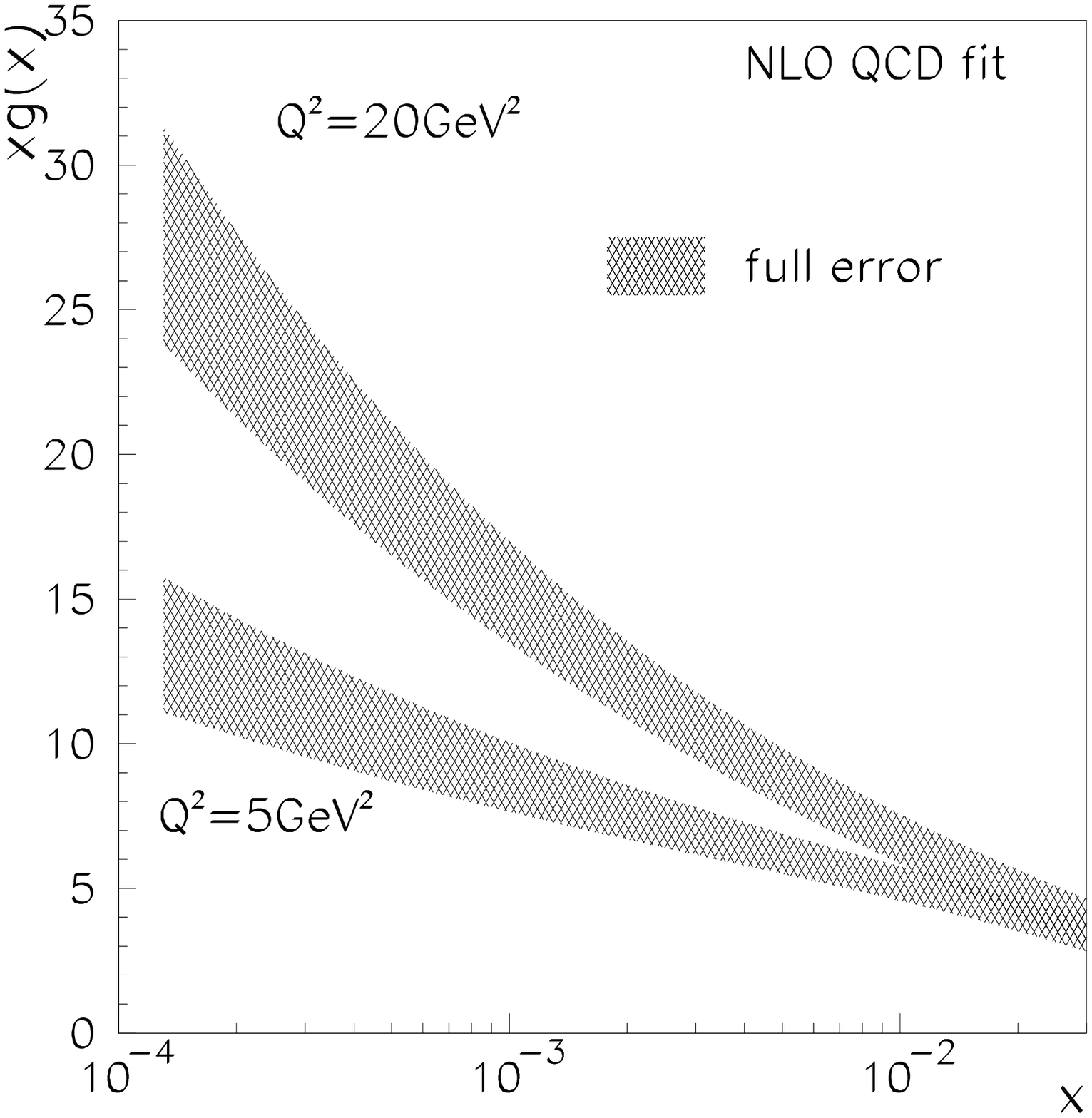,width=5.3cm,
bbllx=-35pt,bblly=145pt,bburx=600pt,bbury=650pt}
\epsfig{file=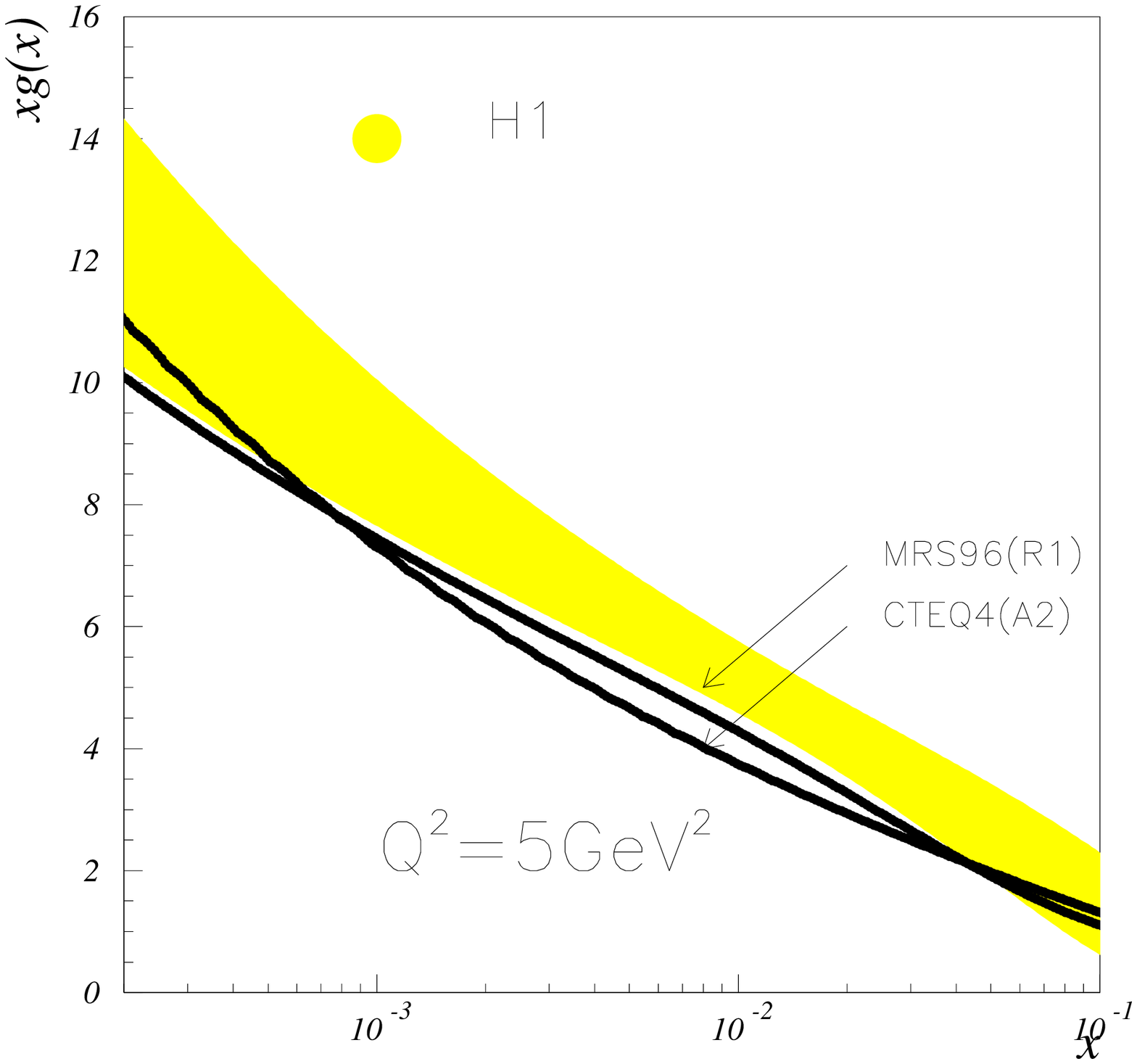,width=5.8cm,
bbllx=15pt,bblly=145pt,bburx=650pt,bbury=720pt}
\end{center}
\caption[]{\label{fig9}
The \MS\ gluon density $xg(x)$ at $Q^2 = $ 5 GeV$^2$ and
$Q^2 = $ 20 GeV$^2$ derived from a NLO
QCD fit of structure function data by H1,\cite{F294H1} and 
the gluon density $xg(x)$ at $Q^2 = $ 5 GeV$^2$
together with results of new global fits from the CTEQ\cite{tung} and 
MRS.\cite{roberts}}
\end{figure}

\subsection{The Gluon}

The gluon distribution at low $x$ is of particular interest, since it 
is this that drives the growth in the quark
distribution. Although many less inclusive 
methods of determining the gluon have been proposed,\cite{mardev}
the analysis of the scaling violations of $F_2$ remains the best 
method at small $x$. Indeed the gluon is now fairly well determined 
since the $F_2$ data are so precise, the Bjorken scaling violations relatively 
large, and driven essentially by the steep rise in the gluon.
The results of some of the various fits are compared in fig.\ref{fig9}.
The H1 fit, based on structure function data only
(for details see ref.~4), also includes error bands
derived from the statistics and experimental systematics. This error bands
have been reduced with more than a factor two compared to the results
based on earlier HERA data. Similar preliminary results became available
from ZEUS.\cite{kooijman} Note that 
the gluon gets steeper as $Q^2$ is increased: as explained above this is 
only possible in perturbative QCD if it rises more slowly than a power of $x$.
 
However, care must be taken as there is no unique way 
of extracting the gluon: besides depending on the choice of scheme and 
the value of $\alpha_s$, it also 
depends to some extent on the assumed parametrization
at the input $Q^2_0$ value and on the treatment of the charm 
threshold.\cite{zomer} Near threshold it seems 
necessary to use the boson gluon fusion mechanism for charm production
rather than charm production via massless evolution.\cite{barone,vogt}
It is also difficult to ascertain the experimental and theoretical 
uncertainties. Recently the HERA experiments have provided 
a full error matrix of the systematic uncertainties in the data, which 
make a full error analysis on the gluon density possible. The two groups 
providing global fits to the data and extracting parton densities are 
strongly encouraged to take this information into account in the future.

\subsection{Calculations of \fc, $F_L$, Direct Photons and Jet Rates}

An equally clean but entirely independent determination of the 
gluon density could in principle be made from an accurate measurement  
of the longitudinal structure function $F_L$. The results presented at 
this meeting\cite{klein} are consistent with the gluon extracted 
from $F_2$ using NLO perturbative QCD (see fig.8), 
albeit with a large systematic error. 

A measurement of the contribution to $F_2$ which arises from 
charm production, \fc\ has also been advocated as a sensitive probe
of the gluon density. It was shown recently\cite{vogt} that
at small $x$ the NLO QCD predictions are stable, with 
scale variations of less than $\pm 10\%$, and that they offer a rather local 
measurement of the gluon. 
The dominant uncertainty in the QCD calculations arises from non-perturbative 
effects conveniently expressed by the uncertainty in the charm quark mass.  
Although the present level of precision of 
the data does not yet allow yet the extraction of the gluon from the \fc\ data,
they are consistent with NLO perturbative calculations using gluons 
obtained from global fits (see fig.9).

The \MS\ gluon distribution may also be determined from the $2+1$ 
jet rate measured at HERA,\cite{rosenbauer,repond} since a full NLO   
perturbative calculation of the partonic cross-section is now 
available.\cite{graudenz,mirkes} There is good agreement with 
gluon distributions found in global fits, the method being particularly 
useful in the intermediate $x$ range ($0.01\lsim x\lsim 0.1$).

Further information on the gluon distribution, especially at larger $x$
($x> 0.01$)
can be extracted from the Tevatron $p\overline{p}$
collider data on direct photon and jet production.
New measurements~\cite{lin} of the production of single isolated photons 
show good 
agreement with NLO QCD calculations for photons with a transverse energy 
larger than 30 GeV. At lower $E_T$ values the data overshoots theory, 
which can possibly be explained by extra $k_T$ generated by initial state 
parton showers. The CDF inclusive 
single jet cross section measurement~\cite{paga}
shows a remarkable excess over theory for transverse momenta larger than
200 GeV. Some doubt has been cast on the assertion\cite{mrsij,roberts} 
that it is not possible to absorb the excess 
by adjusting parton distributions at large $x$: CTEQ have found a 
reasonable (if not optimal) fit to both CDF and BCDMS data,\cite{tung}
if one is ready to accept a somewhat less conventional shape for the
gluon at high $x$. No rise at high-$E_T$ is seen in the D0 data, which 
however have larger systematic errors in this region.
The precise inclusive jet data at lower $E_T$ 
can however be used to give strong constraints on the gluon in the 
intermediate $x$ range, and CTEQ now include this data in some of their global 
fits.\cite{tung}

\begin{figure}[hbt]
\vskip -0.7truecm
\vbox{\hbox{
\hfil\epsfxsize=3.6truecm\epsfbox{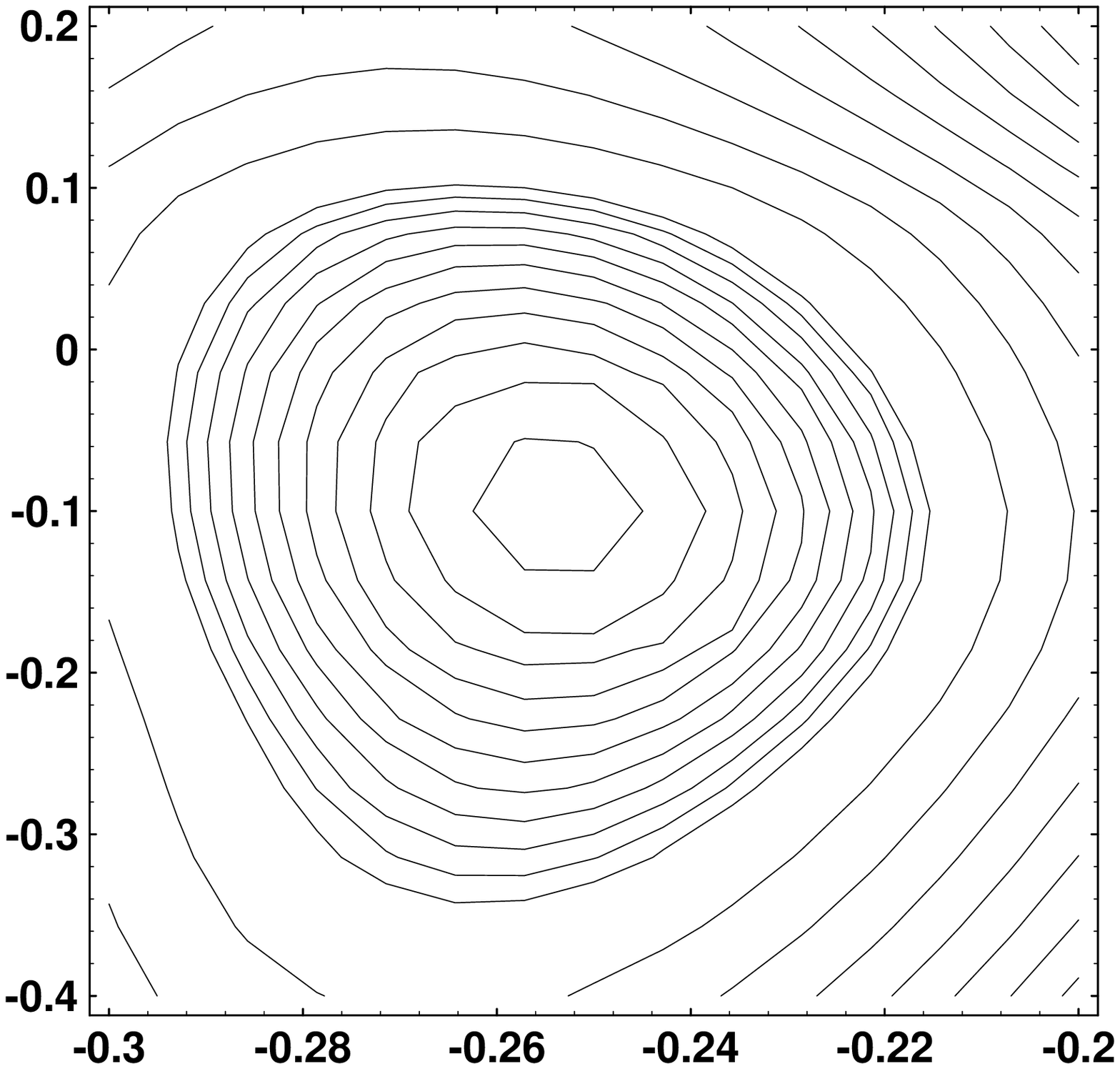}
\hskip 0.3truecm
\epsfxsize=3.6truecm\epsfbox{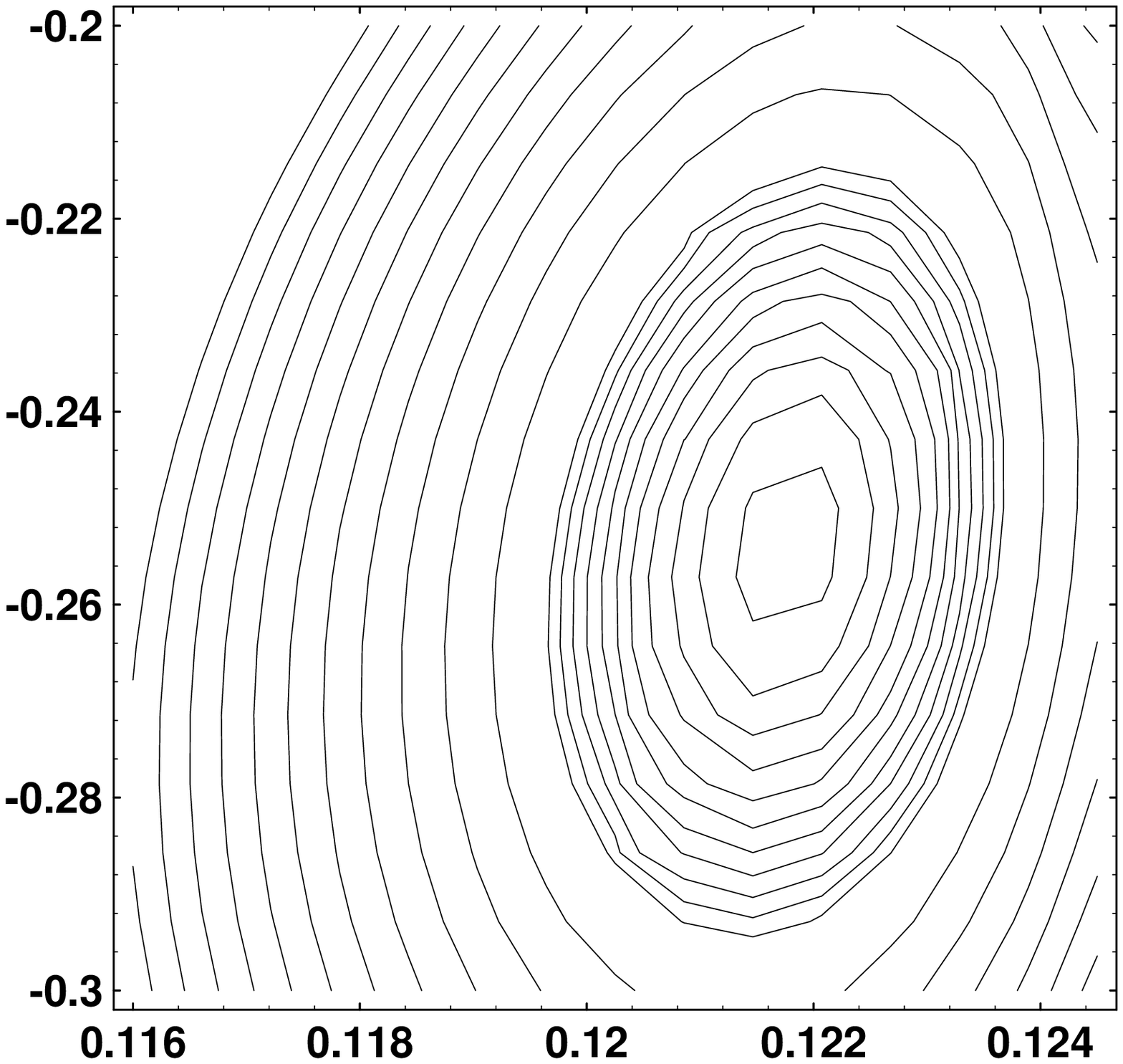}
\hskip 0.3truecm
\epsfxsize=3.6truecm\epsfbox{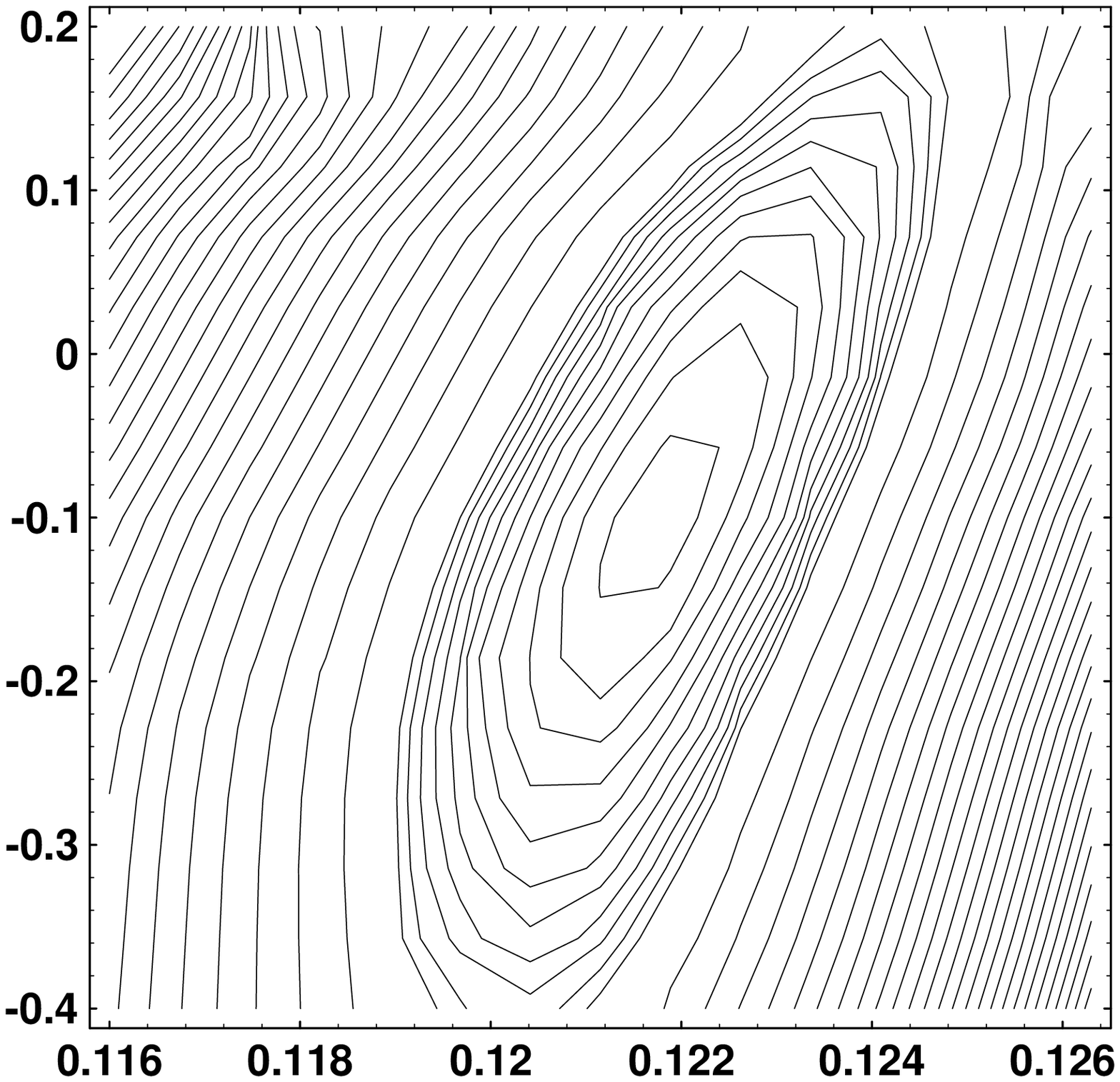}
\hfil}}
\vskip-0.5cm
\caption[]{\label{fig11}
Contour plots of $\chi^2$ in the three orthogonal planes
$(\lambda_S,\lambda_g)$, $(\alpha_s(m_Z),\lambda_S)$ and
$(\alpha_s(m_Z),\lambda_g)$ through the global minimum. The first eleven
contours are at intervals of one unit, while those thereafter are at
intervals of five units.\cite{ball} The data are from H1.\cite{F294H1}}
\end{figure}

\begin{figure}[htbp]
\begin{center}
\epsfxsize=\hsize
\epsfbox{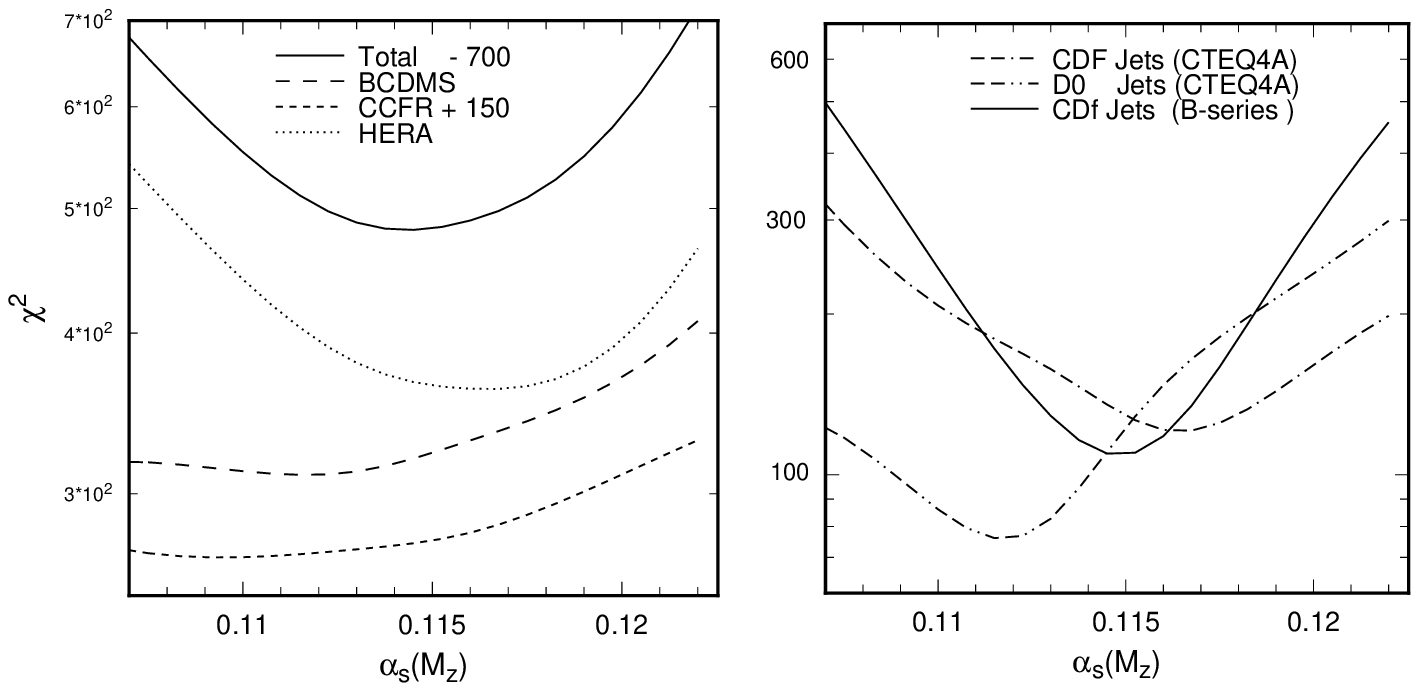}
\end{center}
\vskip-0.5cm
\caption[]{\label{fig12} $\chi^2$ vs. $\alpha_s(M_Z)$ for 
global fits based on 
current experiments: (a) BCDMS, CCFR, combined HERA collider experiments, 
and total of DIS+DY; (b) $\chi^2$ of CDF and D0 jets, calculated using 
statistical errors only,  
using CTEQ4A and B-series parton distributions.\cite{tung}}
\end{figure}

\subsection{The Strong Coupling}

The success of conventional NLO perturbative QCD at small $x$, both 
in accounting for the shape of $F_2$ and in consistently describing 
other observables driven by the gluon distribution, suggests that 
a determination of $\alpha_s$ from HERA data may be possible. Furthermore, 
since the double scaling rise of $F_2$ depends 
strongly (and nonlinearly) on $\alpha_s$, but only weakly on 
the shape of the distributions at large $x$, precise data 
on $F_2$ at small $x$ and large $Q^2$ (say $Q^2\gsim 5$ \GVsq) provide 
an opportunity for a relatively clean measurement.\cite{das} The first such 
determination,\cite{bfas1,bfas2} using data from the 1993 run, gave a 
value
\begin{equation} 
{\alpha_s(M_Z)=0.120\pm 0.005{\rm (exp.)}\pm 0.009{\rm (th.)}:}  
\label{assx}
\end{equation}
this is currently being updated using 1994 data.\cite{ball} The statistical 
significance of the determination may be seen in the $\chi^2$ contour plots 
fig.\ref{fig11}: with the new data, the experimental error is 
considerably reduced. 
Despite some initial scepticism,\cite{mrsas} the result (\ref{assx}) has since
been confirmed in global fits\cite{tung,roberts} 
(see for example fig.\ref{fig12}a). It 
is interesting because it is somewhat larger than the BCDMS result\cite{VM} 
\begin{equation}
{\alpha_s(M_Z)=0.113\pm 0.003{\rm (exp.)}\pm 0.005{\rm (th.)},}
\label{aslx}
\end{equation}
but is instead closer to determinations 
made using time-like processes, such as the $2+1$ jet rate at 
HERA.\cite{tpojets}  

The CDF inclusive jet data with medium-$E_T$ (that is $50\lsim E_T 
\lsim 200$\GV) also seem to prefer\cite{ggy,mrsij} the larger value of 
$\alpha_s$, although similar data from D0 now prefer lower 
values.\cite{tung,roberts} The impact of the jet data on the value 
of $\alpha_s$ in the CTEQ fits may be judged from fig.\ref{fig12}b. 
MRS are now 
providing two sets of fits: R1 with $\alpha_s = 0.113$ and R2 with 
$\alpha_s = 0.120$. R2 provides a significantly better fit to both the 
HERA $F_2$ data and the CDF inclusive jet data, while the fixed target 
structure function data generally prefer R1.

\section{Higher Twists and Higher Logarithms}

Before the advent of HERA data it was widely expected\cite{HERA92} that 
the effects either of higher order logarithms of $1/x$ (summed to give the 
`hard pomeron') or multiparton correlations (leading to `screening' or 
`shadowing') would be very important at small $x$, and invalidate the use of 
conventional NLO perturbative QCD. At this meeting results were presented 
which put quantitative empirical constraints on the size of these effects, 
and explanations were put forward as to why they might be so small.    

\subsection{Higher Twists}

\begin{figure}[htbp]
\begin{center}
\epsfig{figure=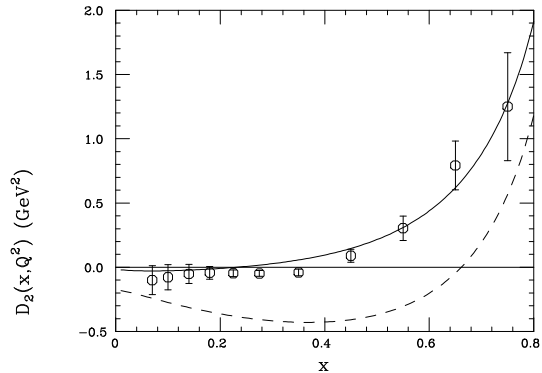,height=5.0cm}
\end{center}
\caption[]{\label{fig13} Higher twist corrections to 
$F_2$ as deduced from BCDMS and SLAC data.\cite{VM}
Also shown are the results of renormalon estimates for higher twist 
corrections to $F_2$ (solid) and $xF_3$ (dashed).\cite{webber}}
\end{figure}

The standard way of quantifying the effects of higher twists in deep 
inelastic cross sections is to 
rerun the perturbative NLO fits to the $F_2$ data using instead
\begin{equation} 
{F_2(x,Q^2)=F_2^{\rm NLO}(x,Q^2)(1+D_2(x)/Q^2):}  
\label{htwist}
\end{equation}
an estimate of the size of the higher twist effects is then given by 
the size of $D_2(x)$ fitted in various bins in $x$. Fits 
to SLAC and BCDMS data show\cite{VM} that at large $x$ $D_2$ is of 
order $1$ \GVsq, but falls rapidly, becoming small and negative for 
$x\lsim 0.3$ (see fig.\ref{fig13}). Recent fits to $F_3$ data follow the same 
pattern,\cite{sidorov} while fits to the HERA $F_2$ data\cite{forte}
give $D_2\simeq 0.2\pm 0.2$\GVsq\ for $10^{-4}\lsim x\lsim10^{-2}$.  

The numbers obtained in all these estimates must be treated 
with caution: twist-4 contributions will not evolve with $Q^2$  
in precisely the same way as the leading twist contribution, and  
leading twist evolution is difficult to compute reliably in the 
vicinity of the charm threshold. However, it 
seems that except at large $x$ the characteristic scale for higher twist 
effects is no more than a few hundred MeV, and may indeed be much smaller. 
This is entirely consistent with theoretical estimates of 
higher twist effects in nonsinglet channels based on 
renormalons.\cite{webber} Such calculations also suggest that twist-6
contributions to structure functions are also very small at small $x$.
In fact at presently obtainable energies no `unitarization' corrections 
are necessary since the inelastic cross-section is always far below any 
reasonable unitarity bound.

Of course at sufficiently small $Q^2$ conventional perturbation theory
must eventually break down for all values of $x$, since in order to 
maintain finiteness of the cross-section $F_2\sim O(Q^2)$, 
$F_L\sim O(Q^4)$ as $Q^2\to 0$. However there is as yet no quantitative 
theory of inelastic lepton-proton scattering at low $Q^2$. 
Fits to low $Q^2$ data\cite{levy} generally have two ingredients: the $x$ 
dependence given by Regge theory (with a pomeron intercept of $1.08$) and 
the $Q^2$ dependence given by a combination of vector meson dominance and 
the photo-production limit. Such parameterizations give a 
satisfactory account of the new ZEUS data
for $Q^2\lsim 1$ \GVsq (see fig.6). The fit to the data 
can be maintained above $1$ \GVsq\ only if the pomeron intercept is allowed to 
rise with $Q^2$: some groups claim to be able to calculate this rise 
non-perturbatively, while others use a parameterization based on the 
rise expected from perturbative QCD.

\subsection{BFKL Logarithms}

At small $x$ one might expect conventional perturbation theory 
to eventually break down, even when $Q^2$ is large enough for 
$\alpha_s(Q^2)$ to be small, due to large logarithms of $1/x$. Naively one 
expects at $n$-loops terms of the form $\alpha_s^n \ln^{2n-1}1/x$,
but for the sufficiently inclusive quantities, such as the singlet 
component of $F_2$, these double logarithms cancel to all orders. Single 
logarithms, of the form $\alpha_s^n \ln^{n}1/x$, remain however.
At fixed coupling they may be summed up using the BFKL 
equation\cite{lipatov,martin} to give a gluon distribution which
rises as $x^{-\lambda}$ in the Regge limit, with 
$\lambda = 12\ln2\alpha_s/\pi \sim 1/2$. However this behaviour  
is far too singular to be compatible with HERA structure function 
data;\cite{das} as explained above, a steep power-like rise would spoil
double scaling. Recent fits\cite{tung,roberts,forte} now deliver a flat 
or valence-like input gluon (see for example fig.\ref{fig3}). 
  
Many explanations have been put forward for the absence of the perturbative 
pomeron in inclusive quantities. In particular it is often 
said that since at present only leading order logarithms can be included, 
the as yet unknown subleading 
effects might somehow reduce the rate of the rise. Another major source of 
uncertainty are the contributions from the small-$k_T$ region; 
the overall normalization of the perturbative rise is rather sensitive 
to the value of an infrared cutoff, and phenomenological studies of the 
perturbative contribution to the total cross-section suggest that it is 
completely swamped by the soft contribution.\cite{landshoff} Nonetheless,
the Lipatov approach, or some development thereof, remains at present 
our only hope of actually being able to compute the behaviour of 
structure functions in the Regge limit. Considerable effort has been put
into numerical solutions of modified BFKL or CCFM equations, with some 
success in obtaining reasonable fits to HERA 
data.\cite{martin,royon}

A less ambitious way of including the higher order logarithms is to 
observe that at large $Q^2$ we can project out the leading logarithmic 
component 
of the BFKL equation,\cite{Jaro} retaining only an infinite (but convergent) 
series of contributions to the perturbative splitting function. Factorization
of infrared singularities is accomplished by the $k_T$-factorization 
theorem,\cite{CH,catani} a generalization of the more usual 
mass-factorization, and likewise proven formally to all orders in perturbation 
theory. The effect of the higher logarithms is then interpreted not as 
a prediction of the form of the input distribution, but as a systematic 
modification of the subsequent perturbative evolution with 
$Q^2$,\cite{EKL}$^{\!-\,}$\cite{EHW} and as 
such can be searched for in the data.
 
There are several reasons why the resulting effects turn out to be    
small:\\
$\bullet$ In the gluon channel the coefficients of the LO logarithms of 
$1/x$ happen to vanish at $O(\alpha_s^2)$, $O(\alpha_s^3)$ and 
$O(\alpha_s^5)$, and the remaining coefficients turn out to be rather 
small. However, although in the quark channel the splitting 
functions only begin at NLO in logarithms of $1/x$, their 
coefficients are relatively large.\cite{CH} Presumably the same is true at NLO 
in the gluon channel, although as yet these subleading terms have not been 
calculated.\\ 
$\bullet$ When $\alpha_s$ runs with $Q^2$, the effect of 
the higher logarithms dies away as $Q^2$ increases,\cite{summing} and is thus 
concentrated in a narrow wedge at small $x$ and low $Q^2$.\\   
$\bullet$ Much of the effect of the higher logarithms may be absorbed in 
the initial condition: this is the essence of factorization. When comparing 
calculations using different perturbative expansions it is thus essential to 
refit the boundary condition: calculations which simply evolve from a 
fixed boundary condition\cite{EHW,blumlein} inevitably over-estimate the 
effect of the higher-order logarithms.\\ 
$\bullet$ Although the LO logarithms are scheme independent, the size 
of the NLO logarithms depends on the choice of factorization scheme.
A universal (i.e. process-independent) reduction in the size of all NLO 
logarithms may be made by absorbing the (singular) gluon normalization 
factor into the input distribution.\cite{ciafaloni} Furthermore, it is possible
to find a scheme in which the NLO logarithms in the quark channel 
vanish.\cite{catanisdis} Alternatively, one can argue for the adoption of a 
`physical' scheme\cite{catani} in which quarks and gluons are directly 
identified with $F_2$ and $F_L$ at small $x$: the NLO logarithms are then
non-zero, but relatively small.\\
$\bullet$ Subleading logarithms may be chosen in various ad hoc ways in order 
to impose momentum conservation by hand.\cite{EHW} In fact it is 
possible to choose a particular 
class of factorization schemes in which momentum is conserved at 
NLO, fixing completely the NLO logarithms in the gluon channel 
in terms of those in the quark channel.\cite{mom} Momentum conservation 
generally reduces the subasymptotic effect of the logarithms, 
since the subleading terms subtract from the leading.\\
$\bullet$ Retaining the higher order logarithms in the large-$x$ region 
is clearly meaningless: splitting functions at more than two loops will 
only be well approximated by their leading $\ln 1/x$ approximation
when $x$ is small. It is thus important to only include the higher 
loop contributions when $x<x_0$, with $x_0$ a suitable (but essentially 
unknown) parameter.
In practice this may be done\cite{summing,bfas2} by replacing 
$\ln\smallfrac{1}{x}$ by $\theta(x_0-x)\ln\smallfrac{x_0}{x}$ in all 
terms beyond two loops, thereby ensuring continuity at $x=x_0$. 
As $x_0\to 0$ the
effect of the higher order logarithms is then reduced systematically to zero: 
the empirically determined value of $x_0$ thus gives a useful 
parameterization of the 
constraints put on the size of the logarithms by the HERA data. 

\begin{figure*}[t]
\begin{center}\vskip-3.5truecm
\mbox{\hskip-.6truecm
\epsfig{figure=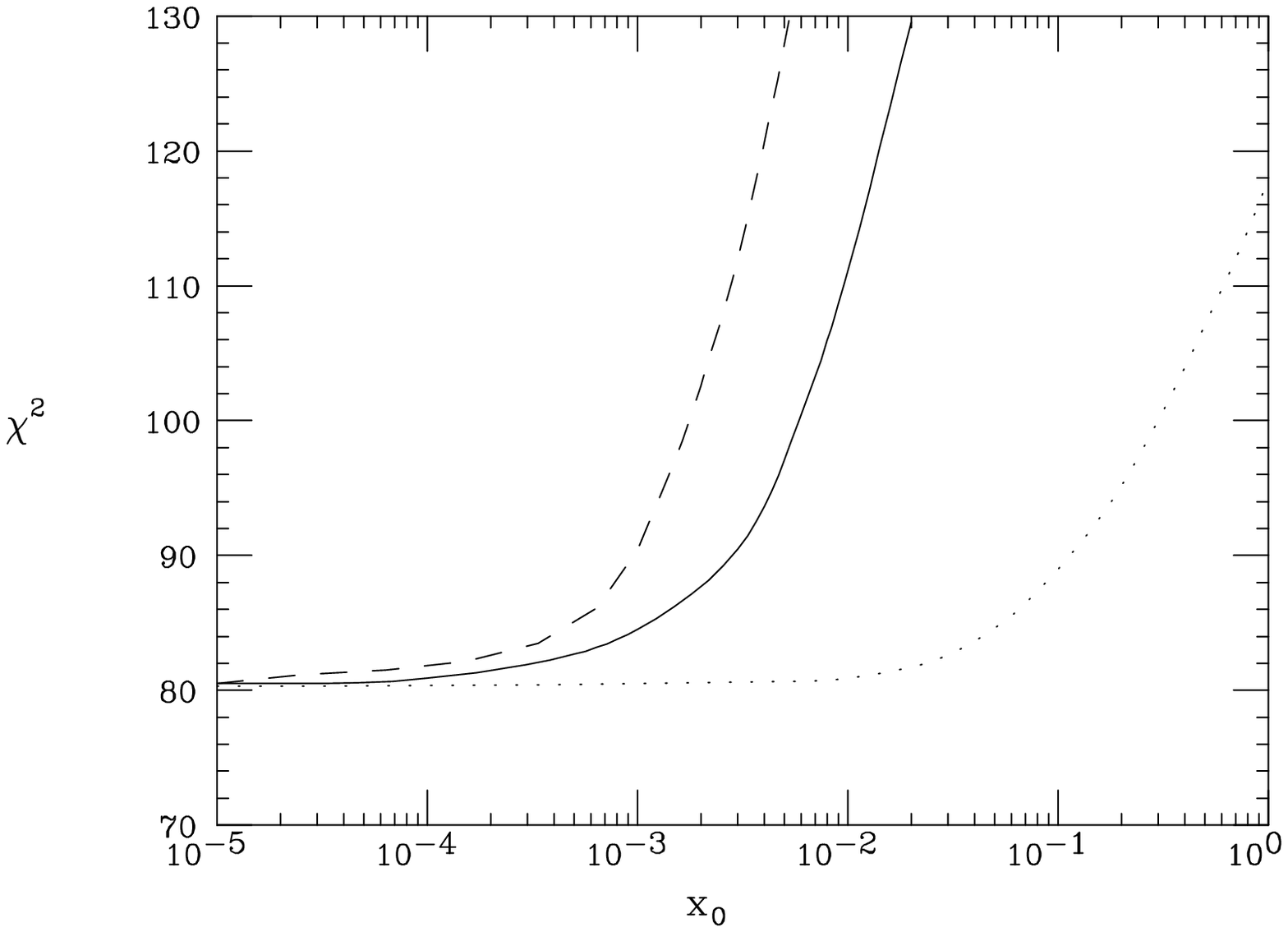,height=8.5truecm}
\epsfig{figure=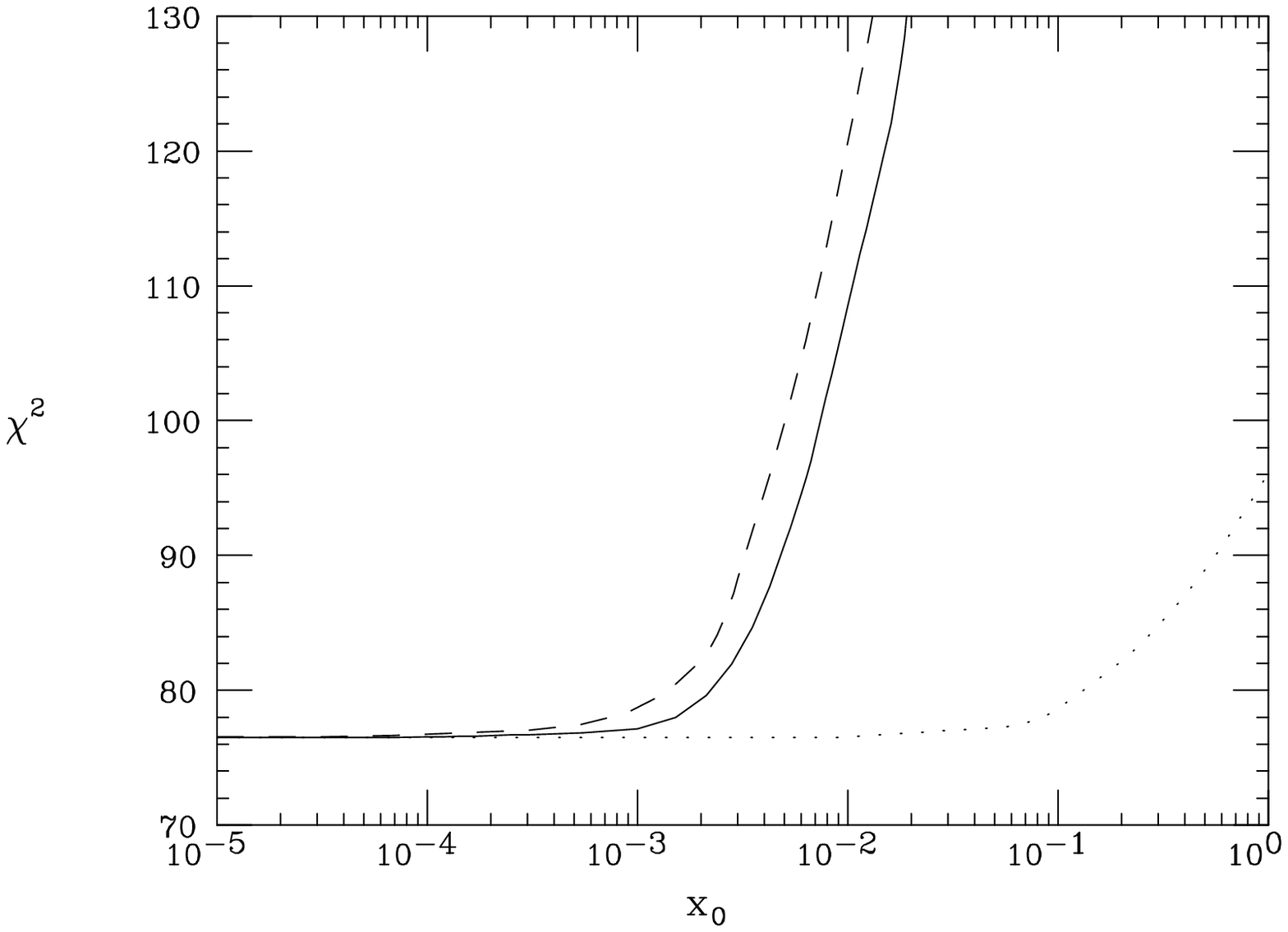,height=8.5truecm}}
\vskip-2.5truecm
\caption[]{\label{fig14}
The $\chi^2$ of the fits to the H1 1994 data as a function of $x_0$ for 
a variety of factorization schemes.\cite{forte} Those in the 
left-hand figure are all \MS\ schemes, while those in the right-hand figure 
are DIS schemes. The dotted curves are two loop calculations with LO 
BFKL logarithms added, while the 
dashed and solid lines are complete NLO order calculations in the schemes of 
refs.62,66\ 
respectively.
The boundary condition is refitted at each value of $x_0$, and all the NLO 
schemes conserve momentum.}
\end{center}\end{figure*}

The first calculations including higher-order logarithms in which all these
considerations were properly taken into account\cite{bfas1,bfas2} 
found that with the 1993 HERA data excellent fits could be achieved 
at NLO with $x_0=0.1$ in all schemes, while values of $x_0$ as large as unity
were acceptable only when LO logarithms are summed, or in NLO schemes where 
the NLO logarithms are suppressed.
However it was also noted\cite{scheme} that there were large scheme dependent 
fluctuations at small $x$ and low $Q^2$. Since the 1994 data, besides being 
much more precise than the 1993 data, also explore this region, they are much
more sensitive to the effects of the higher-order logarithms. It turns out 
that the new data effectively exclude such logarithms: in 
each scheme the best fit to the data is obtained as $x_0\to 0$, and in 
the complete NLO calculations a reasonable fit is only 
obtained when $x_0$ is unnaturally small (see fig.\ref{fig14}). 

It should also be noted that the NLO logarithms in the quark channel would
lead to a suppression of the gluon distribution extracted from $F_2$ at 
small $x$ by a factor of around two, and thus to a corresponding 
suppression of both $F_L$ and \fc\ as compared to the two loop 
expectation.\cite{CH}$^{\!-\,}$\cite{EHW,FRT,thorne} No such suppression 
is seen in the data, which if anything lie a little above the two 
loop predictions (see figs.\ref{fig7} and~\ref{fig8}, also ref.72).

In our opinion a completely satisfactory explanation for the apparent 
absence of these contributions to perturbative evolution at small $x$ 
has not yet been put forward.

\subsection{Other Logarithms}

The cancellation of the double logarithms in the singlet component of $F_2$,
$F_L$ and \fc\ does not occur for either nonsinglet components of $F_2$ 
or indeed for nonsinglet or singlet contributions to the 
polarized structure function  $g_1$.\cite{KL,ermolaev,webber} 
When the double logarithms are naively incorporated into the 
evolution equations in the same way as the single logarithms 
were above,\cite{ermolaev,blumlein} large effects at 
small $x$ can be generated dynamically from flat or valence-like 
inputs: for example $g_1^p$ grows so rapidly that its first moment 
apparently diverges. However it would be premature to take 
these predictions at face value. As yet there is no $k_T$-factorization 
theorem when the double logarithms are uncancelled, and it is not 
even clear whether it might eventually be possible to prove one.  
A better understanding of double logarithms will probably require 
first a true understanding of single logarithms.

In this connection, it should also be noted that in many less 
inclusive quantities, and in particular observables related to jets, 
the double logarithms are also uncancelled.\cite{marchesini} 

All the logarithms discussed so far are ultraviolet logarithms, in the 
sense that they only become important at high energies. At lower 
energies infrared logarithms can be more important. This applies 
particularly to the evolution of structure functions at large $x$, 
where it becomes necessary to consider terms of the form 
$\alpha_s^n\ln^n(1-x)$. Such terms might be important in the determination 
of $\alpha_s$ from fixed target data,\cite{VM} in the interpretation of
high-$E_T$ inclusive jet rate as measured by CDF (through the 
comparison with evolved fixed target data),\cite{tung} and in the 
evolution of the pomeron structure function (since the gluon distribution 
in the pomeron\cite{newman} seems to be very strongly peaked at 
large $\beta$).  

Consider the theoretical error quoted in the determinations (\ref{assx}) and 
(\ref{aslx}) of $\alpha_s$. In both cases the dominant contribution 
to the theoretical errors
comes from a crude estimate of NNLO corrections based on a variation of the 
renormalization and factorization scales in the range $0.25Q^2<\mu^2<4Q^2$. 
The error in the small $x$ determination (\ref{assx}) also includes 
an error from the effect of the logarithms of $1/x$ (though we now know 
empirically that the effect of such logarithms is very small): the 
corresponding determination of the error in the large $x$ result 
(\ref{aslx}) due to infrared logarithms has yet to be made. 

There are several indications that the effects of infrared 
logarithms on the analysis of fixed target data at large $x$ may be 
much more greater than is usually supposed. Firstly,
when the $W^2$ cut on the BCDMS data is raised from $10$ \GVsq\ to $30$ \GVsq, 
the value of $\alpha_s$ increases substantially.\cite{roberts} Secondly,
for first moment sum rules such as the Gross--Llewellyn-Smith sum rule,
which obtain their dominant contribution from the large $x$ region, the NNLO 
correction is known to be as large as $35\%$ of the NLO, while the NNNLO 
correction is as large as $15\%$.\cite{GLSSR} All of these corrections 
have the same sign. Thirdly, higher twist corrections, which
in the renormalon approach are considered to be directly related to the 
size of higher order corrections,\cite{webber} are strongly peaked at 
large $x$ (see fig.\ref{fig13}). 

A calculation of the three loop splitting function
would go some way towards settling this issue: partial results in 
the nonsinglet channel, which may prove useful for error estimation, have 
been presented recently.\cite{LRV}

\section{Conclusions}

There has been considerable progress in both the range and precision of 
structure function data over the last year: in particular 
the 1994 data on $F_2$ are now final, there are new data at low $Q^2$ and $x$,
and new measurements of $R$ and \fc. These data have reinforced the 
main conclusion of the Paris meeting, namely that conventional NLO 
perturbative QCD works remarkably well for $Q^2\gsim 1$\GVsq, matching 
smoothly onto Regge expectations below this value. The gluon distribution 
is now determined with an uncertainty of around $10\%$ for $x$ down to 
$10^{-4}$.

However there is now definite, though admittedly inconclusive, evidence 
that collider structure function data prefer a higher 
value of the strong coupling than fixed target data, more in 
keeping with high energy determinations from $e^+e^-$ machines.
Moreover the tight constraints the new data place on both higher 
twists and, perhaps more surprisingly, on higher logarithms of $1/x$ 
at small $x$ suggest that perturbative evolution at high energies 
is still not yet completely understood. 
The new data are a continuing inspiration to theorists searching for 
a deeper understanding of perturbative QCD.

\section*{Acknowledgments}
We would like to thank all the contributors to our working group
for their presentations and participation in the 
lively and stimulating discussions, Robin Devenish for
co-organizing this session, Stefano Forte for a careful reading of 
the final manuscript, and Giulio D'Agostini and his team 
for their excellent organization of the conference.

\end{document}